\documentclass{article}

\usepackage{amsmath,latexsym,amssymb}
\usepackage[pdftex]{graphicx}
\usepackage[colorlinks=true]{hyperref}
\usepackage{blkarray}
\usepackage{geometry}
\newcommand{\diag}{\ensuremath{\mathrm{diag}}}
\newcommand{\Rank}{\ensuremath{\mathrm{Rank}}}
\newcommand{\smacol}[2]{\ensuremath{\bigl(\begin{smallmatrix} #1 \\ #2 \end{smallmatrix}\bigr)}}
\newcommand{\smamat}[4]{\ensuremath{\bigl(\begin{smallmatrix} #1 & #2 \\ #3 & #4 \end{smallmatrix}\bigr)}}
\newtheorem{theorem}{Theorem}

\begin{document}

\title{\Large{\textbf{Implementation of Bilinear Hamiltonian Interactions between Linear Quantum Stochastic Systems via Feedback}}}

\author{Symeon Grivopoulos$^*$ \and Ian R. Petersen
\thanks{The authors are with the School of Engineering and Information Technology, UNSW Canberra, Canberra BC 2610, Australia, {\tt\small symeon.grivopoulos@gmail.com, i.r.petersen@gmail.com}. This work was supported by the Australian Research Council under grant FL110100020}}

\maketitle

\begin{abstract}
A number of recent works employ bilinear Hamiltonian interactions between Linear Quantum Stochastic Systems (LQSSs). Contrary to naturally occurring Hamiltonian interactions between physical systems, such interactions must be engineered. In this work, we propose a simple model for the implementation of an arbitrary bilinear interaction between two given LQSSs via a feedback interconnection.
\end{abstract}


\section{Introduction}
\label{sec:Introduction}

Linear Quantum Stochastic Systems (LQSSs) are a class of models used in quantum optics \cite{garzol00,walmil08,wismil10}, circuit QED systems \cite{matjirper11, kerandku13}, quantum opto-mechanical systems \cite{tsacav10, masheipir11, hammab12, donfiokuz12}, and elsewhere. The mathematical framework for these models is provided by the theory of quantum Wiener processes, and the associated Quantum Stochastic Differential Equations \cite{par99,mey95,hudpar84}. Potential applications of LQSSs include quantum information processing, quantum measurement and control. In particular, an important application of LQSSs is as coherent quantum feedback controllers for other quantum systems, i.e. controllers that do not perform any measurement on the controlled quantum system, and thus, have the potential to outperform classical controllers, see e.g. \cite{yankim03a,yankim03b,jamnurpet08,nurjampet09,maapet11b,zhajam12,mab08,hammab12,critezsoh13}.

The ways LQSSs can interact are of particular importance to applications such as the synthesis of larger LQSSs in terms of simple ones, the design of coherent quantum observers and controllers for LQSSs, etc. There is, of course, the usual directional signal connection from the output of one system to the input of another. This sort of coupling of LQSSs is referred to as indirect, or field-mediated interaction, and, depending on the sort of connection, namely feedforward or feedback, it can be uni- or bi-directional. Such interconnections have been considered in \cite{jamnurpet08,nurjampet09}, for example, in the context of coherent quantum controller synthesis, and in \cite{nurjamdoh09,nur10b,nur10a,nurgripet16,gripet15} in the context of synthesis of LQSSs. Additionally, we may have a direct or Hamiltonian interaction between the LQSSs. This sort of coupling, which results from physical interaction between quantum systems, is bidirectional. In this work, a direct or Hamiltonian interaction between the LQSSs is always meant to be bilinear, see Subsection \ref{subsec:Bidirectional Field-Mediated and Hamiltonian Interactions of Linear Quantum Stochastic Systems}. Such interactions have been considered in \cite{nurjamdoh09,zhajam11,pet14,miajamugr15, sicvlapet15} for the applications mentioned above.

Contrary to the Hamiltonian interactions between physical systems, like atoms or subatomic particles, that occur naturally, such interactions between engineered LQSSs must themselves be engineered. In \cite[Subsection 6.4]{nurjamdoh09}, a scheme for the implementation of a direct interaction between two one degree-of-freedom LQSSs (generalized harmonic oscillators) is proposed. However, this implementation becomes rather involved if one wants to create more complicated direct interactions between larger dimensional LQSSs. In \cite{pet15,pet16}, an implementation is proposed for the coherent quantum observer of \cite{pet14} directly coupled to a one degree-of-freedom LQSSs, both with and without input/output channels for the observer. The approach taken in these works is, to construct the composite plant + observer system. A drawback of this approach is that it would not be applicable in situations where the ``individuality'' of the two LQSSs must be preserved.


In this paper, we propose a new method for the implementation of an arbitrary bilinear interaction between two given LQSSs of any dimensions, via feedback. Our method entails modifying the original LQSSs, by adding input/output ports, and modifying their self-Hamiltonians, see Theorem \ref{thm:DIs via FMIs}. Since the interacting LQSSs can have an arbitrary number of degrees-of-freedom and inputs/outputs, the proposed model is not described in the detail of the constructions in \cite{nurjamdoh09,pet15, pet16}. However, the modified LQSSs and the static linear network necessary for the implementation of the direct interaction, see Section \ref{sec:Bilinear Hamiltonian Interactions via Feedback}, can be implemented using the general synthesis results in \cite{nurjamdoh09,nur10a,reczeiber94,bra05, leoneu04}. Such a general method would be useful, among other things, in implementing coherent quantum controllers that employ direct interaction with the plant, such as those considered in \cite{zhajam11, sicvlapet15}. Also, even though the method is proposed in the context of LQSSs, it can be applied to quantum stochastic systems with non-linear dynamics, as well, as the dynamics of the systems play no part in the implementation of the Hamiltonian interaction. Indeed, the later is a result of feedback through additional (linear) inputs/outputs created in the two systems.

The rest of the paper is organized, as follows: In Section \ref{sec:Background Material}, we establish some notation and terminology used in the paper, and provide a short overview of LQSSs and direct/indirect couplings between them. In Section \ref{sec:Bilinear Hamiltonian Interactions via Feedback}, we present our model for the implementation of an arbitrary bilinear interaction between two given LQSSs via field-mediated ones, see Theorem \ref{thm:DIs via FMIs}, and prove its validity. Section \ref{sec:An Illustrative Example} contains an illustrative example.

\section{Background Material}
\label{sec:Background Material}

\subsection{Notation and terminology}
\label{subsec:Notation and terminology}

We begin by establishing notation and terminology that will be used throughout this paper:
\begin{enumerate}
  \item For $x \in \mathbb{R}$, $\lceil x \rceil$ is the smallest integer greater or equal to $x$. $x^{*}$ denotes the complex conjugate of a complex number $x$ or the adjoint of an operator $x$, respectively. For a matrix $X=[x_{ij}]$ with number or operator entries, $X^{\#}=[x_{ij}^*]$, $X^{\top}=[x_{ji}]$ is the usual transpose, and $X^{\dag}=(X^{\#})^{\top}$. The commutator of two operators $X$ and $Y$ is defined as $[X,Y]=XY-YX$.
  \item The identity matrix in $n$ dimensions will be denoted by $I_n$, and a $r \times s$ matrix of zeros will be denoted by $0_{r \times s}$. Let $\mathbb{J}_{2k}=\bigl(\begin{smallmatrix} 0_{k \times k} & I_k \\ -I_k & 0_{k \times k} \end{smallmatrix}\bigr)$. When the dimensions can be inferred from context, we shall simply use $I$, $\mathbf{0}$, and $\mathbb{J}$. $\delta_{ij}$ denotes the Kronecker delta symbol, i.e. $I=[\delta_{ij}]$. Also, $\left(\begin{smallmatrix} X_1 \\ X_2 \\ \vdots \\ X_k \\ \end{smallmatrix}\right)$ is the vertical concatenation of the matrices $X_1,X_2,\ldots,X_k$, of equal column dimension, $(\,Y_1 \, Y_2 \, \ldots \, Y_k \,)$ is the horizontal concatenation of the matrices $Y_1, Y_2, \ldots, Y_k$ of equal row dimension, and $\diag(Z_1,Z_2,\ldots,Z_k)$ is the block-diagonal matrix formed by the square matrices $Z_1,Z_2,\ldots,Z_k$.
  \item For a $2r \times 2s$ matrix $X$, define its $\sharp$-\emph{adjoint} $X^{\sharp}$, by $X^{\sharp}= -\mathbb{J}_{2s}X^{\dag}\mathbb{J}_{2r}$. The $\sharp$-\emph{adjoint} satisfies properties similar to the usual adjoint, namely $(x_1 A + x_2 B)^{\sharp}=x_1^* A^{\sharp} + x_2^* B^{\sharp}$, $(AB)^{\sharp}=B^{\sharp}  A^{\sharp}$, and $(A^{\sharp})^{\sharp}=A$.
  \item A $2k \times 2k$ complex matrix T is called \emph{symplectic}, if it satisfies $TT^{\sharp}=T^{\sharp}T=I_{2k} \Leftrightarrow  T \mathbb{J}_{2k} T^{\dag}=T^{\dag}\mathbb{J}_{2k} T=\mathbb{J}_{2k}$. Hence, any symplectic matrix is invertible, and its inverse is its $\sharp$-adjoint. The set of these matrices forms a non-compact Lie group known as the symplectic group. Real symplectic matrices constitute a subgroup of the (complex) symplectic group.
\end{enumerate}

\subsection{Linear Quantum Stochastic Systems}
\label{subsec:Linear Quantum Stochastic Systems}

The material in this subsection is fairly standard, and our presentation aims mostly at establishing notation and terminology. To this end, we follow the review paper \cite{pet10}. For the mathematical background necessary for a precise discussion of LQSSs, some standard references are \cite{par99,mey95,hudpar84}, while for a Physics perspective, see \cite{garzol00,garcol85}. The references \cite{nurjamdoh09,edwbel05,goujam09,gougohyan08,goujamnur10} contain a lot of relevant material, as well.

The systems we consider in this work are collections of quantum harmonic oscillators interacting among themselves, as well as with their environment. The $i$-th harmonic oscillator ($i=1,\ldots,n$) is described by its position and momentum variables, $q_i$ and $p_i$, respectively. These are self-adjoint operators satisfying the \emph{Canonical Commutation Relations} (CCRs) $[q_i,q_j]=0$, $[p_i,p_j]=0$, and $[q_i,p_j]=\imath\delta_{ij}$, for $i,j=1,\ldots,n$.  If we define the vectors of operators $q=(q_1,q_2,\dots,q_n)^{\top}$, $p=(p_1,p_2,\ldots,p_n)^{\top}$, and $x=\bigl(\begin{smallmatrix} q \\ p \end{smallmatrix} \bigr)$, the CCRs can be expressed as
\begin{eqnarray}
[x,x^{\top}] \doteq x x^{\top} -(x x^{\top})^{\top} =
\left(\begin{array}{cc}
\mathbf{0} & \imath I_n \\
-\imath I_n & \mathbf{0} \\
\end{array}\right)
= \imath \mathbb{J}_{2n}. \label{eq:Real CCRs}
\end{eqnarray}

The environment is modelled as a collection of bosonic heat reservoirs. The $i$-th heat reservoir ($i=1,\ldots,m$) is described by bosonic \emph{field annihilation and creation operators} $\mathcal{A}_i(t)$ and $\mathcal{A}_i^*(t)$, respectively. The field operators are \emph{adapted quantum stochastic processes} with forward differentials $d\mathcal{A}_i(t)= \mathcal{A}_i(t+dt)-\mathcal{A}_i(t)$, and $d\mathcal{A}_i^*(t)= \mathcal{A}_i^*(t+dt)-\mathcal{A}_i^*(t)$. They satisfy the quantum It\^{o} products $d\mathcal{A}_i(t) d\mathcal{A}_j(t)=0$, $d\mathcal{A}_i^*(t) d\mathcal{A}_j^*(t)=0$, $d\mathcal{A}_i^*(t) d\mathcal{A}_j(t)=0$, and $d\mathcal{A}_i(t) d\mathcal{A}_j^*(t)=\delta_{ij} dt$. If we define the vector of field operators $\mathcal{A}(t)=(\mathcal{A}_1(t),\mathcal{A}_2(t),\dots,\mathcal{A}_m(t))^{\top}$, and the vector of self-adjoint field quadratures
\[ \mathcal{V}(t)=\frac{1}{\sqrt{2}}
\left(\begin{array}{c}
\mathcal{A}(t)+\mathcal{A}(t)^{\#} \\
\imath(\mathcal{A}(t)-\mathcal{A}(t)^{\#}) \\
\end{array}\right), \]
the quantum It\^{o} products above can be expressed as
\setlength{\arraycolsep}{1pt}
\begin{eqnarray}
d\mathcal{V}(t) d\mathcal{V}(t)^{\top}=\frac{1}{2} \left(\!\!\begin{array}{cc}
I_m & \imath I_m \\
-\imath I_m & I_m \\
\end{array}\!\!\right) dt= \frac{1}{2}\,(I_{2m} + \imath \mathbb{J}_{2m}) dt. \label{eq:Real Quantum Ito products}
\end{eqnarray}
\setlength{\arraycolsep}{3pt}

To describe the dynamics of the harmonic oscillators and the quantum fields, we introduce certain operators. We begin with the Hamiltonian operator $H=\frac{1}{2}x^{\top}Rx$, which specifies the dynamics of the harmonic oscillators in the absence of any environmental influence. $R^{2n \times 2n}$ is a real symmetric matrix referred to as the Hamiltonian matrix. Next, we have the coupling operator $L$ (vector of operators) that specifies the interaction of the harmonic oscillators with the quantum fields. $L$ depends linearly on the position and momentum operators of the oscillators, and can be expressed  as $L=L_q q + L_p p$. We construct the real coupling matrix $C^{2m \times 2n}$ from $L_q^{m \times n}$ and $L_p^{m \times n}$, as 
\[ C= \left(\begin{array}{cc}
L_q + L_q^{\#} & L_p + L_p^{\#} \\ 
-\imath(L_q - L_q^{\#}) & -\imath(L_p - L_p^{\#})
\end{array}\right) . \]
Finally, we have the unitary scattering matrix $S^{m \times m}$, that describes the interactions between the quantum fields themselves.

In the \emph{Heisenberg picture} of quantum mechanics, the joint evolution of the harmonic oscillators and the quantum fields is described by the following system of \emph{Quantum Stochastic Differential Equations} (QSDEs):
\begin{eqnarray}
dx &=& (\mathbb{J} R -\frac{1}{2}C^{\sharp}C) x dt -C^{\sharp}D\, d\mathcal{V}, \nonumber \\
d\mathcal{V}_{out}&=& C x dt + D\, d\mathcal{V}, \label{eq:Real General LQSS}
\end{eqnarray}
where
\[ D = \frac{1}{2} \left(\begin{array}{cc}
S+S^{\#} & \imath (S-S^{\#}) \\
-\imath (S-S^{\#}) & S+S^{\#} \\
\end{array}\right), \]
is a $2m \times 2m$ real orthogonal symplectic matrix. The field quadrature operators $\mathcal{V}_{i \, out}(t)$ describe the outputs of the system. (\ref{eq:Real General LQSS}) is a description of the dynamics of the LQSS in the so-called \emph{real quadrature operator representation}, where the states, inputs, and outputs are all self-adjoint operators. We are going to use a version of (\ref{eq:Real General LQSS}) generalized in two ways: First, we replace the real orthogonal symplectic transformation $D$,  with a more general real symplectic transformation $D$, see e.g. \cite{goujamnur10} for a discussion of this in the context of the  \emph{creation-annihilation representation}. Second, in the context of coherent quantum systems in particular, the output of a quantum system may be fed into another quantum system, so we substitute the more general input and output notations $\mathcal{U}$ and $\mathcal{Y}$, for $\mathcal{V}$ and $\mathcal{V}_{out}$, respectively. The resulting QSDEs are the following:
\begin{eqnarray}
dx &=& (\mathbb{J} R -\frac{1}{2}C^{\sharp}C) x dt -C^{\sharp}D\, d\mathcal{U}, \nonumber \\
d\mathcal{Y}&=& C x dt + D\, d\mathcal{U}, \label{eq:Real General LQSS 2}
\end{eqnarray}
The forward differentials $d\mathcal{U}$ and $d\mathcal{Y}$ of inputs and outputs, respectively (or, more precisely, of their quadratures), contain ``quantum noises'', as well as a ``signal part'' (linear combinations of variables of other systems). One can prove that, the structure of  (\ref{eq:Real General LQSS 2}) is preserved under linear transformations of state $\bar{x}= T x$, if and only if $T$ is real symplectic. From the point of view of quantum mechanics, $T$ must be real symplectic so that the transformed position and momentum operators are also self-adjoint and satisfy the same CCRs, as one can verify from (\ref{eq:Real CCRs}).

In Subsection \ref{subsec:Bidirectional Field-Mediated and Hamiltonian Interactions of Linear Quantum Stochastic Systems}, we shall need a description of a LQSS with its inputs/outputs partitioned into two groups. Let the $m$-dimensional vector of input fields $\mathcal{A}(t)$ be partitioned in blocks of dimension $m_a$ and $m_b$, respectively ($m=m_a + m_b$), as follows:
\begin{eqnarray*}
\mathcal{A}(t)=
\left(\begin{array}{c}
\mathcal{A}_1(t) \\
\mathcal{A}_2(t) \\
\vdots \\
\mathcal{A}_m(t) \\
\end{array}\right)=
\left(\begin{array}{c}
\mathcal{A}_a(t) \\
\mathcal{A}_b(t) \\
\end{array}\right).
\end{eqnarray*}
For the vectors of input field quadratures of the two groups of inputs,
\begin{eqnarray*}
\mathcal{V}_a(t)=\frac{1}{\sqrt{2}}
\left(\begin{array}{c}
\mathcal{A}_a(t)+\mathcal{A}_a(t)^{\#} \\
\imath(\mathcal{A}_a(t)-\mathcal{A}_a(t)^{\#}) \\
\end{array}\right), \ \mathrm{and} \
\mathcal{V}_b(t)=\frac{1}{\sqrt{2}}
\left(\begin{array}{c}
\mathcal{A}_b(t)+\mathcal{A}_b(t)^{\#} \\
\imath(\mathcal{A}_b(t)-\mathcal{A}_b(t)^{\#}) \\
\end{array}\right),
\end{eqnarray*}
respectively, we have that $\smacol{\mathcal{V}_a(t)}{\mathcal{V}_b(t)} = \Pi \mathcal{V}(t)$, where
\begin{eqnarray}
\Pi =
\begin{blockarray}{ccccc}
m_a & m_b & m_a & m_b & \\
\begin{block}{(cccc)c}
I_{m_a} & 0 & 0 & 0 & m_a \\
0 & 0 & I_{m_a} & 0 & m_a \\
0 & I_{m_b} & 0 & 0 & m_b \\
0 & 0 & 0 & I_{m_b} & m_b \\
\end{block}
\end{blockarray} . \label{eq:Definition of Pi}
\end{eqnarray}
$\Pi$ is a $2m \times 2m$ real orthogonal matrix, hence $\Pi^{-1}=\Pi^{\top}$. We have the analogous relation $\smacol{\mathcal{V}_{a \, out}(t)}{\mathcal{V}_{b \, out}(t)} = \Pi \mathcal{V}_{out}(t)$, for the corresponding output field quadratures. If we define
\begin{eqnarray*}
\hat{C} &=&
\left(\begin{array}{c}
\hat{C}_a \\
\hat{C}_b \\
\end{array}\right)= \Pi C, \\
\hat{D} &=&
\left(\begin{array}{cc}
\hat{D}_{aa} & \hat{D}_{ab} \\
\hat{D}_{ba} & \hat{D}_{bb} \\
  \end{array}\right) = \Pi D \Pi^{\top},
\end{eqnarray*}
the second equation of (\ref{eq:Real General LQSS}) becomes
\begin{eqnarray*}
\left(\begin{array}{c}
d\mathcal{V}_{a \, out} \\
d\mathcal{V}_{b \, out} \\
\end{array}\right) =\hat{C}x dt + \hat{D}
\left(\begin{array}{c}
d\mathcal{V}_a \\
d\mathcal{V}_b \\
\end{array}\right) =
\left(\begin{array}{c}
\hat{C}_a x dt + \hat{D}_{aa} d\mathcal{V}_a + \hat{D}_{ab} d\mathcal{V}_b \\
\hat{C}_b x dt + \hat{D}_{ba} d\mathcal{V}_a + \hat{D}_{bb} d\mathcal{V}_b \\
\end{array}\right) .
\end{eqnarray*}
Finally, using the identity $\Pi \mathbb{J}_{2m} \Pi^{\top}=\diag(\mathbb{J}_{2m_a},\mathbb{J}_{2m_b})$, the first equation of (\ref{eq:Real General LQSS}) takes the form
\begin{eqnarray*}
dx = (\mathbb{J} R -\frac{1}{2}\hat{C}_a^{\sharp}\hat{C}_a -\frac{1}{2}\hat{C}_b^{\sharp}\hat{C}_b) x dt
- (\hat{C}_a^{\sharp}\hat{D}_{aa} + \hat{C}_b^{\sharp}\hat{D}_{ba}) d\mathcal{V}_a
   -(\hat{C}_a^{\sharp}\hat{D}_{ab} + \hat{C}_b^{\sharp}\hat{D}_{bb}) d\mathcal{V}_b .
\end{eqnarray*}
Putting everything together, and employing the more general notation $(\mathcal{U}_a,\mathcal{Y}_a)$, $(\mathcal{U}_b,\mathcal{Y}_b)$ for the inputs/outputs of each group, we have the following description:
\begin{eqnarray}
dx &=& (\mathbb{J} R -\frac{1}{2}\hat{C}_a^{\sharp}\hat{C}_a -\frac{1}{2}\hat{C}_b^{\sharp}\hat{C}_b) x dt
- (\hat{C}_a^{\sharp}\hat{D}_{aa} + \hat{C}_b^{\sharp}\hat{D}_{ba}) d\mathcal{U}_a
   -(\hat{C}_a^{\sharp}\hat{D}_{ab} + \hat{C}_b^{\sharp}\hat{D}_{bb}) d\mathcal{U}_b , \nonumber \\
d\mathcal{Y}_a &=& \hat{C}_a x dt + \hat{D}_{aa} d\mathcal{U}_a + \hat{D}_{ab} d\mathcal{U}_b, \nonumber \\
d\mathcal{Y}_b &=& \hat{C}_b x dt + \hat{D}_{ba} d\mathcal{U}_a + \hat{D}_{bb} d\mathcal{U}_b . \label{eq:Real General LQSS partitioned i/o}
\end{eqnarray}

\subsection{Bidirectional Field-Mediated and Hamiltonian Interactions of Linear Quantum Stochastic Systems}
\label{subsec:Bidirectional Field-Mediated and Hamiltonian Interactions of Linear Quantum Stochastic Systems}

In this subsection, we review bidirectional field-mediated and Hamiltonian interactions between LQSSs.

\begin{figure}[!h]
\begin{center}
\scalebox{.4}{\includegraphics{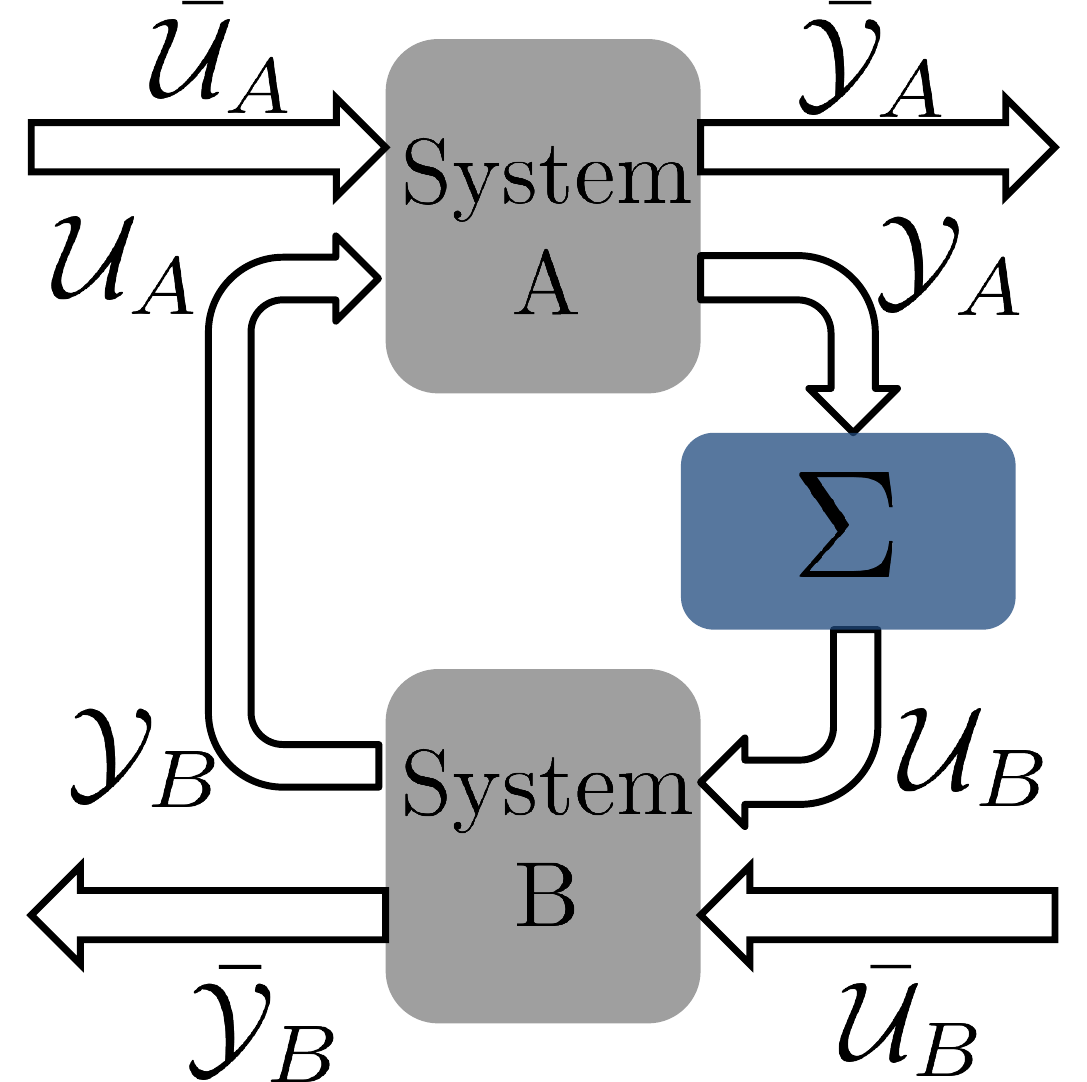}} \caption{Graphical representation of a bidirectional field-mediated interaction between LQSSs $A$ and $B$.} \label{fig:FMI}
\end{center}
\end{figure}

Let $A$ be a LQSS with $n_A$ number of modes, Hamiltonian matrix $R_A$, and two groups of inputs/outputs: $m_A$ inputs/outputs $(\bar{\mathcal{U}}_A, \bar{\mathcal{Y}}_A)$ with coupling matrix $\bar{C}_A \in \mathbb{R}^{2m_A \times 2n_A}$, and $m$ inputs/outputs $(\mathcal{U}_A,\mathcal{Y}_A)$ with coupling matrix $C_A \in \mathbb{R}^{2m \times 2n_A}$. The QSDEs for $A$ are the following:
\begin{eqnarray*}
dx_A &=& [\mathbb{J} R_A -\frac{1}{2}\bar{C}_A^{\sharp}\bar{C}_A -\frac{1}{2}C_A^{\sharp}C_A]\, x_A dt \\
&-& (\bar{C}_A^{\sharp}\bar{D}_A + C_A^{\sharp} D_{A\,21}) d\bar{\mathcal{U}}_A - (\bar{C}_A^{\sharp} D_{A\,12} + C_A^{\sharp} D_A) d\mathcal{U}_A,  \\
d\bar{\mathcal{Y}}_A &=& \bar{C}_A x_A dt + \bar{D}_A d\bar{\mathcal{U}}_A + D_{A\,12}d\mathcal{U}_A, \\
d\mathcal{Y}_A &=& C_A x_A dt + D_{A\,21} d\bar{\mathcal{U}}_A + D_A d\mathcal{U}_A,
\end{eqnarray*}
where $\bar{D}_A$, $D_{A\, 12}$, $D_{A\, 21}$, and $D_A$ are, respectively, $2m_A \times 2m_A$, $2m_A \times 2m$, $2m \times 2m_A$, and $2m \times 2m$ real matrices such that $\Pi_A^{\top} \smamat{\bar{D}_A}{D_{A\, 12}}{D_{A\, 21}}{D_A} \Pi_A$ is real symplectic, and $\Pi_A$ is defined as in (\ref{eq:Definition of Pi}), with $m_a=m_A$, and $m_b=m$. Let, also, $B$ be a LQSS with $n_B$ number of modes, Hamiltonian matrix $R_B$, and two groups of inputs/outputs: $m_B$ inputs/outputs $(\bar{\mathcal{U}}_B, \bar{\mathcal{Y}}_B)$ with coupling matrix $\bar{C}_B \in \mathbb{R}^{2m_B \times 2n_B}$, and $m$ inputs/outputs $(\mathcal{U}_B,\mathcal{Y}_B)$ with coupling matrix $C_B \in \mathbb{R}^{2m \times 2n_B}$. The QSDEs for $B$ are the following:
\begin{eqnarray*}
dx_B &=& [\mathbb{J} R_B -\frac{1}{2}\bar{C}_B^{\sharp}\bar{C}_B -\frac{1}{2}C_B^{\sharp}C_B]\, x_B dt  \\
&-& (\bar{C}_B^{\sharp}\bar{D}_B + C_B^{\sharp} D_{B\,21}) d\bar{\mathcal{U}}_B - (\bar{C}_B^{\sharp} D_{B\,12} + C_B^{\sharp} D_B) d\mathcal{U}_B,  \\
d\bar{\mathcal{Y}}_B &=& \bar{C}_B x_B dt + \bar{D}_B d\bar{\mathcal{U}}_B + D_{B\,12}d\mathcal{U}_B,  \\
d\mathcal{Y}_B &=& C_B x_B dt + D_{B\,21} d\bar{\mathcal{U}}_B + D_B d\mathcal{U}_B,
\end{eqnarray*}
where $\bar{D}_B$, $D_{B\, 12}$, $D_{B\, 21}$, and $D_B$ are, respectively, $2m_B \times 2m_B$, $2m_B \times 2m$, $2m \times 2m_B$, and $2m \times 2m$ real matrices such that $\Pi_B^{\top} \smamat{\bar{D}_B}{D_{B\, 12}}{D_{B\, 21}}{D_B} \Pi_B$ is real symplectic, and $\Pi_B$ is defined as in (\ref{eq:Definition of Pi}), with $m_a=m_B$, and $m_b=m$. A bidirectional indirect, or field-mediated interaction between them, is defined by the feedback interconnection conditions $\mathcal{U}_B = \Sigma \mathcal{Y}_A$, and $\mathcal{U}_A = \mathcal{Y}_B$, where $\Sigma$ is a $2m \times 2m$ real symplectic matrix, see Figure \ref{fig:FMI}.
The above model of a LQSSs bidirectional interaction via feedback is a fairly general one. For our purposes, it suffices to consider a simpler model with $D_{A\, 12}=0$, $D_{A\, 21}=0$, $D_A=I$, and correspondingly for system $B$. The QSDEs for the simplified model are the following:
\begin{eqnarray}
dx_A &=& [\mathbb{J} R_A -\frac{1}{2}\bar{C}_A^{\sharp}\bar{C}_A -\frac{1}{2}C_A^{\sharp}C_A]\, x_A dt - \bar{C}_A^{\sharp}\bar{D}_A d\bar{\mathcal{U}}_A - C_A^{\sharp} d\mathcal{U}_A, \label{eq:QSDE system A FMI}  \\
dx_B &=& [\mathbb{J} R_B -\frac{1}{2}\bar{C}_B^{\sharp}\bar{C}_B -\frac{1}{2}C_B^{\sharp}C_B]\, x_B dt - \bar{C}_B^{\sharp}\bar{D}_B d\bar{\mathcal{U}}_B - C_B^{\sharp} d\mathcal{U}_B, \label{eq:QSDE system B FMI}  \\
d\bar{\mathcal{Y}}_A &=& \bar{C}_A x_A dt + \bar{D}_A d\bar{\mathcal{U}}_A, \label{eq:QSDE system A FMI 3} \\
d\bar{\mathcal{Y}}_B &=& \bar{C}_B x_B dt + \bar{D}_B d\bar{\mathcal{U}}_B, \label{eq:QSDE system B FMI 3} \\
d\mathcal{Y}_A &=& C_A x_A dt + d\mathcal{U}_A, \label{eq:QSDE system A FMI 2} \\
d\mathcal{Y}_B &=& C_B x_B dt + d\mathcal{U}_B, \label{eq:QSDE system B FMI 2} \\
\mathcal{U}_B &=& \Sigma \mathcal{Y}_A, \  \mathcal{U}_A = \mathcal{Y}_B. \label{eq: FMI Interconnection conditions}
\end{eqnarray}

\begin{figure}[!h]
\begin{center}
\scalebox{.4}{\includegraphics{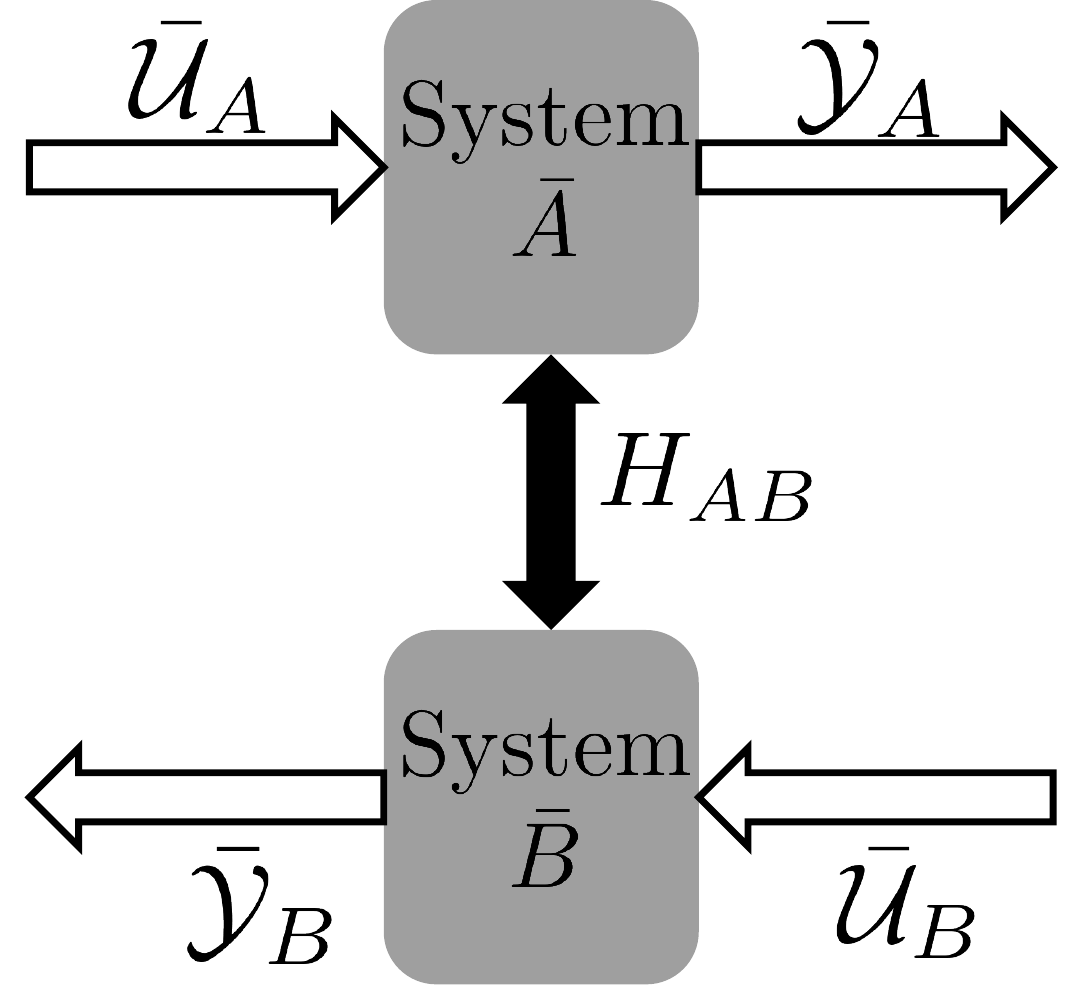}}
\caption{Graphical representation of a Hamiltonian interaction between LQSSs $\bar{A}$ and $\bar{B}$.} \label{fig:DI}
\end{center}
\end{figure}

Let $\bar{A}$ and $\bar{B}$ be two LQSSs with, respectively, $n_A$ and $n_B$ number of modes, $m_A$ and $m_B$ inputs/outputs, and parameters $(\bar{R}_A, \bar{C}_A, \bar{D}_A)$ and $(\bar{R}_B, \bar{C}_B, \bar{D}_B)$. A direct, or Hamiltonian interaction between them, given by the interaction Hamiltonian $H_{AB}= \bar{x}_A^{\top} \bar{R}_{AB}\bar{x}_B$, where $\bar{R}_{AB}$ is a real $2n_A \times 2n_B$ matrix, defines the composite system $\overline{AB}$ with $(n_A + n_B)$ number of modes, $(m_A + m_B)$ inputs/outputs, and Hamiltonian matrix $\left( \begin{smallmatrix} \bar{R}_A & \bar{R}_{AB} \\ \bar{R}_{AB}^{\top} & \bar{R}_B \end{smallmatrix} \right)$. The QSDES for the composite system, are the following:
\begin{eqnarray}
d\bar{x}_A &=& [\mathbb{J} \bar{R}_A  -\frac{1}{2}\bar{C}_A^{\sharp}\bar{C}_A]\, \bar{x}_A dt + \mathbb{J} \bar{R}_{AB}\bar{x}_B dt - \bar{C}_A^{\sharp} \bar{D}_A d\bar{\mathcal{U}}_A, \label{eq:QSDE system A DI} \\
d\bar{x}_B &=& [\mathbb{J} \bar{R}_B -\frac{1}{2}\bar{C}_B^{\sharp}\bar{C}_B]\, \bar{x}_B dt + \mathbb{J} \bar{R}_{AB}^{\top} \bar{x}_A dt - \bar{C}_B^{\sharp} \bar{D}_B d\bar{\mathcal{U}}_B, \label{eq:QSDE system B DI} \\
d\bar{\mathcal{Y}}_A &=& \bar{C}_A \bar{x}_A dt + \bar{D}_A d\bar{\mathcal{U}}_A, \label{eq:QSDE system A DI 2} \\
d\bar{\mathcal{Y}}_B &=& \bar{C}_B \bar{x}_B dt + \bar{D}_B d\bar{\mathcal{U}}_B. \label{eq:QSDE system B DI 2}
\end{eqnarray}
A graphical representation of the composite LQSS $\bar{A}\bar{B}$ is given in Figure \ref{fig:DI}.

\section{Bilinear Hamiltonian Interactions via Feedback}
\label{sec:Bilinear Hamiltonian Interactions via Feedback}

In this section, we prove that an arbitrary bilinear Hamiltonian interaction between two LQSSs $\bar{A}$ and $\bar{B}$, see (\ref{eq:QSDE system A DI}) - (\ref{eq:QSDE system B DI 2}), can be realized by the feedback interconnection (\ref{eq:QSDE system A FMI}) - (\ref{eq: FMI Interconnection conditions}) of two LQSSs $A$ and $B$ , with related parameters. We have the following result:

\begin{theorem}\label{thm:DIs via FMIs}
The model (\ref{eq:QSDE system A DI}) - (\ref{eq:QSDE system B DI 2}) of the LQSSs $\bar{A}$ and $\bar{B}$ interacting directly via a Hamiltonian interaction, can be realized by the model (\ref{eq:QSDE system A FMI}) - (\ref{eq: FMI Interconnection conditions}) of the field-mediated bidirectional interaction between the LQSSs $A$ and $B$, for any $\bar{R}_{AB} \in \mathbb{R}^{2\bar{n}_A \times 2\bar{n}_B}$. The parameters of the two realizations are related, as follows: We have that, for any $\bar{R}_{AB}$, there exist real $2m \times 2n_A$, and $2m \times 2n_B$ matrices $C_A$ and $C_B$, respectively, and a real $2m \times 2m$ $\sharp$-skew-symmetric matrix $X$ ($X^{\sharp}=-X$), satisfying the relation
\begin{eqnarray}
\bar{R}_{AB} &=& \frac{1}{2}\mathbb{J} C_A^{\sharp}(X+I)C_B, \label{eq:DI/FMI parameter relation}
\end{eqnarray}
as long as $m \geq \lceil\frac{1}{2} \Rank(\bar{R}_{AB})\rceil$. Then,
\begin{eqnarray}
R_A &=& \bar{R}_A - \frac{1}{2}\mathbb{J} \big(C_A^{\sharp} X C_A\big), \label{eq:New Hamiltonian system A} \\
R_B &=& \bar{R}_B - \frac{1}{2}\mathbb{J} \big(C_B^{\sharp} X C_B\big),\ \mathrm{and} \label{eq:New Hamiltonian system B} \\
\Sigma &=& (X-I)(X+I)^{-1}. \label{eq:Static Interconnection Gain}
\end{eqnarray}
\end{theorem}
\textbf{Proof:} We start from the model of two LQSSs interacting via bosonic input-output channels defined by (\ref{eq:QSDE system A FMI})-(\ref{eq: FMI Interconnection conditions}), and show that it reduces to the model of two directly interacting LQSSs described by (\ref{eq:QSDE system A DI})-(\ref{eq:QSDE system B DI}), with the corresponding parameters related by (\ref{eq:DI/FMI parameter relation})-(\ref{eq:Static Interconnection Gain}). For now, the number of interconnection channels $m$ is unspecified. To begin, from (\ref{eq:QSDE system A FMI 2}), (\ref{eq:QSDE system B FMI 2}) and (\ref{eq: FMI Interconnection conditions}), we obtain the equations
\begin{eqnarray*}
d\mathcal{U}_B &=& \Sigma C_A x_A dt + \Sigma d\mathcal{U}_A, \\
d\mathcal{U}_A &=& C_B x_B dt +  d\mathcal{U}_B,
\end{eqnarray*}
which can be cast in the following form:
\begin{eqnarray*}
\left(\begin{array}{rr}
I & -I \\
-\Sigma  & I \\
\end{array}\right)
\left(\begin{array}{c}
d\mathcal{U}_A \\
d\mathcal{U}_B \\
\end{array}\right)=
\left(\begin{array}{c}
C_B x_B \\
\Sigma C_A x_A \\
\end{array}\right) dt.
\end{eqnarray*}
A unique solution exists when $\Sigma$ has no unit eigenvalues, and is given by
\begin{eqnarray*}
\left(\begin{array}{c}
d\mathcal{U}_A \\
d\mathcal{U}_B \\
\end{array}\right) &=&
\left(\begin{array}{ll}
(I-\Sigma)^{-1} & (I-\Sigma)^{-1} \\
(I-\Sigma)^{-1}\Sigma  & (I-\Sigma)^{-1} \\
\end{array}\right)
\left(\begin{array}{c}
C_B x_B \\
\Sigma C_A x_A \\
\end{array}\right) dt \\
&=& \left(\begin{array}{c}
(I-\Sigma)^{-1}[C_B x_B + \Sigma C_A x_A]\,dt \\
(I-\Sigma)^{-1}\Sigma[C_B x_B + C_A x_A]\,dt \\
\end{array}\right).
\end{eqnarray*}
Inserting the expressions for $d\mathcal{U}_A$ and $d\mathcal{U}_B$ from above into (\ref{eq:QSDE system A FMI}) and (\ref{eq:QSDE system B FMI}), respectively, we obtain the following QSDEs:
\begin{eqnarray}
dx_A &=& [\mathbb{J} R_A -\frac{1}{2}\bar{C}_A^{\sharp}\bar{C}_A -\frac{1}{2}C_A^{\sharp}C_A -C_A^{\sharp}(I-\Sigma)^{-1} \Sigma C_A]\, x_A dt \nonumber \\
&-&C_A^{\sharp}(I-\Sigma)^{-1}C_B x_B dt  - \bar{C}_A^{\sharp} \bar{D}_A d\bar{\mathcal{U}}_A, \label{QSDE system A DI1}  \\
dx_B &=& [\mathbb{J} R_B -\frac{1}{2}\bar{C}_B^{\sharp}\bar{C}_B -\frac{1}{2}C_B^{\sharp}C_B -C_B^{\sharp}(I-\Sigma)^{-1}\Sigma C_B]\, x_B dt \nonumber \\
&-&C_B^{\sharp}(I-\Sigma)^{-1} \Sigma C_A x_A dt  - \bar{C}_B^{\sharp} \bar{D}_B d\bar{\mathcal{U}}_B. \label{QSDE system B DI1}
\end{eqnarray}
Now, we introduce the Cayley transform $X=(I+\Sigma)(I-\Sigma)^{-1}$, defined for matrices $\Sigma$ with no unit eigenvalues. Its unique inverse is given by $\Sigma=(X-I)(X+I)^{-1}$. It is straightforward to verify that $X$ is real and $\sharp$-skew-symmetric ($X^{\sharp}=-X$) if and only if $\Sigma$ is real symplectic. Using the identities $(I-\Sigma)^{-1}\Sigma= \frac{1}{2}(X-I)$, and $(I-\Sigma)^{-1}= \frac{1}{2}(X+I)$, (\ref{QSDE system A DI1}) and (\ref{QSDE system B DI1}) take the following form:
\begin{eqnarray}
dx_A &=& [\mathbb{J} R_A -\frac{1}{2}C_A^{\sharp} X C_A -\frac{1}{2}\bar{C}_A^{\sharp}\bar{C}_A]\, x_A dt -
\frac{1}{2}C_A^{\sharp}(X+I)C_B x_B dt  - \bar{C}_A^{\sharp} \bar{D}_A d\bar{\mathcal{U}}_A, \label{QSDE system A DI2}  \\
dx_B &=& [\mathbb{J} R_B -\frac{1}{2}C_B^{\sharp} X C_B -\frac{1}{2}\bar{C}_B^{\sharp}\bar{C}_B]\, x_B dt -
\frac{1}{2}C_B^{\sharp}(X-I)C_A x_A dt  - \bar{C}_B^{\sharp} \bar{D}_B d\bar{\mathcal{U}}_B. \label{QSDE system B DI2}
\end{eqnarray}
We define
\begin{eqnarray*}
\bar{R}_A &=& R_A +\frac{1}{2}\mathbb{J} \big(C_A^{\sharp} X C_A\big), \\
\bar{R}_B &=& R_B +\frac{1}{2}\mathbb{J} \big(C_B^{\sharp} X C_B\big), \\
\bar{R}_{AB} &=& \frac{1}{2}\mathbb{J} C_A^{\sharp}(X+I)C_B.
\end{eqnarray*}
It is straightforward to verify that $\bar{R}_A$ and $\bar{R}_B$ are real symmetric. Also, we have that
\begin{eqnarray*}
\bar{R}_{AB}^{\top} &=&\frac{1}{2}\Big(\mathbb{J} C_A^{\sharp}(X+I)C_B \Big)^{\top} = -\frac{1}{2}\mathbb{J} \Big(\mathbb{J} C_A^{\sharp}(X+I)C_B \Big)^{\sharp} \mathbb{J} \\
&=& -\frac{1}{2}\mathbb{J}\Big( C_B^{\sharp}(X^{\sharp}+I)C_A (-\mathbb{J}) \Big) \mathbb{J} = \frac{1}{2}\mathbb{J} C_B^{\sharp}(X-I)C_A.
\end{eqnarray*}
Using the definitions of $\bar{R}_A$, $\bar{R}_B$, and $\bar{R}_{AB}$, equations (\ref{QSDE system A DI2}) and (\ref{QSDE system B DI2}) simplify to
\begin{eqnarray*}
dx_A &=& [\mathbb{J} \bar{R}_A  -\frac{1}{2}\bar{C}_A^{\sharp}\bar{C}_A]\, x_A dt +
\mathbb{J} \bar{R}_{AB}x_B dt  - \bar{C}_A^{\sharp} \bar{D}_A d\bar{\mathcal{U}}_A, \\
dx_B &=& [\mathbb{J} \bar{R}_B -\frac{1}{2}\bar{C}_B^{\sharp}\bar{C}_B]\, x_B dt +
\mathbb{J} \bar{R}_{AB}^{\top} x_A dt  - \bar{C}_B^{\sharp} \bar{D}_B d\bar{\mathcal{U}}_B,
\end{eqnarray*}
which are exactly (\ref{eq:QSDE system A DI}) and (\ref{eq:QSDE system B DI}), with the corresponding parameters related by (\ref{eq:DI/FMI parameter relation})-(\ref{eq:Static Interconnection Gain}). To complete the proof, we must show that given any real $2n_A \times 2n_B$ matrix $\bar{R}_{AB}$, there exist real $2m \times 2n_A$ and $2m \times 2n_B$ matrices $C_A$ and $C_B$, respectively, and a real $2m \times 2m$ $\sharp$-skew-symmetric matrix $X$ satisfying (\ref{eq:DI/FMI parameter relation}), as long as $m \geq \lceil\frac{1}{2} \Rank(\bar{R}_{AB})\rceil$.

Using the definition of the $\sharp$-adjoint, (\ref{eq:DI/FMI parameter relation}) takes the form
\begin{equation}\label{DI/FMI parameter relation 2}
2 \bar{R}_{AB} = C_A^{\top}(\mathbb{J} X+\mathbb{J})C_B = C_A^{\top}(Y+\mathbb{J})C_B,
\end{equation}
for $Y=\mathbb{J} X$. It is straightforward to show that $Y$ is symmetric, due to $X$ being $\sharp$-skew-symmetric. To proceed, we shall need to define a special form of SVD for matrices with even dimensions. Given a complex matrix $T^{2r \times 2s}$, let $T=U \hat{T}V^{\dag}$ be its usual SVD. $\hat{T}^{2r \times 2s}$ has non-zero elements (the singular values of $T$) only on the main diagonal, i.e. it has one of the following structures, depending on whether $r$ is no greater or no smaller than $s$:
\[
\hat{T} = \begin{blockarray}{ccccccccccc}
&&s&&&&s&&&& \\
\begin{block}{(ccccc|ccccc)c}
*&&&&&&&&&& \\
&*&&&&&&&&& r \\
&&\ddots&&&&&&&& \\ \cline{1-10}
&&&*&&&&&&& \\
&&&&*&&&&&& r \\
&&&&&\ddots&&&&& \\
\end{block}
\end{blockarray},\ \mathrm{or} \
\hat{T} = \begin{blockarray}{ccccccc}
&s&&&s&& \\
\begin{block}{(ccc|ccc)c}
*&&&&&& \\
&*&&&&& \\
&&\ddots&&&& r \\
&&&*&&& \\
&&&&*&& \\ \cline{1-6}
&&&&&\ddots& \\
&&&&&& \\
&&&&&& r \\
&&&&&& \\
&&&&&& \\
\end{block}
\end{blockarray}.
\]
 A number of the last elements on the main diagonal may be zero, depending on the nullity of $T$. Let $\Pi_1$ and $\Pi_2$ be two permutation matrices such that $\tilde{T}=\Pi_1 \hat{T}\Pi_2$ has one of the following structures, depending on whether $r$ is no greater or no smaller than $s$:
\[
\tilde{T} = \begin{blockarray}{ccccccccccc}
&&s&&&&s&&&& \\
\begin{block}{(ccccc|ccccc)c}
*&&&&&&&&&& \\
&*&&&&&&&&& r \\
&&\ddots&&&&&&&& \\ \cline{1-10}
&&&&&*&&&&& \\
&&&&&&*&&&& r \\
&&&&&&&\ddots&&& \\
\end{block}
\end{blockarray},\ \mathrm{or} \
\tilde{T} = \begin{blockarray}{ccccccc}
&s&&&s&& \\
\begin{block}{(ccc|ccc)c}
*&&&&&& \\
&*&&&&& \\
&&\ddots&&&& r \\
&&&&&& \\
&&&&&& \\ \cline{1-6}
&&&*&&& \\
&&&&*&& \\
&&&&&\ddots& r \\
&&&&&& \\
&&&&&& \\
\end{block}
\end{blockarray},
\]
In the case $r \leq s$, this can be achieved with $\Pi_1=I$, and in the case $r \geq s$, it can be achieved with $\Pi_2=I$. However, a non-trivial permutation $\Pi_1$ may be needed in the first case, and a non-trivial permutation $\Pi_2$ in the second case, in order to redistribute the diagonal zeros of $\tilde{T}$. Indeed, let  $\tilde{T}=\bigl(\begin{smallmatrix}\tilde{T}_1 & 0 \\ 0 & \tilde{T}_2 \end{smallmatrix}\bigr)$. Then, $\tilde{T}_1^{r \times s}$ and $\tilde{T}_2^{r \times s}$ have non-zero elements only on the main diagonal. If $T$ has nullity $p$, then we define $\tilde{T}_1$ and $\tilde{T}_2$ to have $p/2$ zeros as their last diagonal entries, for even $p$, and $(p+1)/2$ and $(p-1)/2$ zeros as their last diagonal entries, respectively, for odd $p$. Hence, by construction, $\tilde{T}_1$ and $\tilde{T}_2$ have at most $\lceil\frac{\Rank(T)}{2}\rceil$ non-zero elements on the main diagonal. Then, we can write $T=U \hat{T}V^{\dag}=(U\Pi_1^{\top})\, (\Pi_1\hat{T}\Pi_2)(V\Pi_2)^{\dag}=\tilde{U}\tilde{T}\tilde{V}^{\dag}$, where $\tilde{U}$ and $\tilde{V}$ are column permutations of $U$ and $V$, respectively, hence unitary or real orthogonal, depending on whether $T$ is complex or real.

Let $\bar{R}_{AB}=Q_1 \tilde{R} Q_2^{\top}$ be the special form of the SVD of $\bar{R}_{AB}$ as defined above, and $Y=P^{\top} \tilde{Y} P$ be the eigen-decomposition of $Y$. Notice that $Q_1$, $Q_2$, and $P$ are all real orthogonal matrices, since $\bar{R}_{AB}$ is real and $Y$ is real symmetric. Employing these factorizations in (\ref{DI/FMI parameter relation 2}), results in the following equation:
\begin{eqnarray}
2 Q_1 \tilde{R} Q_2^{\top} &=& C_A^{\top} (P^{\top} \tilde{Y} P + \mathbb{J}\,)\, C_B \Rightarrow \nonumber \\
2 \tilde{R} &=& (P C_A Q_1)^{\top} (\tilde{Y} + P\,\mathbb{J}\,P^{\top}) \, (P C_B Q_2) \Leftrightarrow  \nonumber \\
2 \tilde{R} &=& G_A^{\top} (\tilde{Y} + P\,\mathbb{J}\,P^{\top}) \, G_{B}, \label{DI/FMI parameter relation 3}
\end{eqnarray}
with the obvious definitions $G_A=P C_A Q_1$, and $G_B=P C_B Q_2$. $G_A$ and $G_B$ are real $2m \times 2n_A$, and $2m \times 2n_B$ matrices, respectively. Up to this point, $Y$ was a completely arbitrary $2m \times 2m$ real symmetric matrix, hence $P$ was an arbitrary $2m \times 2m$ real orthogonal matrix. Now, we constrain $P$ to be symplectic as well, namely to satisfy $P P^{\sharp}=I_{2m} \Leftrightarrow P\,\mathbb{J}_{2m}P^{\top}=\mathbb{J}_{2m}$. We make this choice in order to facilitate the construction of solutions of (\ref{DI/FMI parameter relation 3}), and, equivalently, through all the matrix definitions and factorizations, of (\ref{eq:DI/FMI parameter relation}). Making this choice means that, we shall not construct the most general solution to (\ref{DI/FMI parameter relation 3}). However, it will be easier to settle the question of existence of its solutions, and there will remain plenty of free parameters in the proposed solution. Using the fact that $P$ is symplectic, (\ref{DI/FMI parameter relation 3}) simplifies to
\begin{eqnarray}
2 \tilde{R} = G_A^{\top} (\tilde{Y} + \mathbb{J}\,) \, G_{B}. \label{DI/FMI parameter relation 4}
\end{eqnarray}
Now we let $\tilde{R} = \bigl(\begin{smallmatrix}\tilde{R}_1 & 0 \\ 0 & \tilde{R}_2 \end{smallmatrix}\bigr)$, $\tilde{Y} = \bigl(\begin{smallmatrix}\tilde{Y}_1 & 0 \\ 0 & \tilde{Y}_2 \end{smallmatrix}\bigr)$, $G_A = \bigl(\begin{smallmatrix}G_{A1} & G_{A3} \\ G_{A4} & G_{A2} \end{smallmatrix}\bigr)$, and $G_B = \bigl(\begin{smallmatrix}G_{B1} & G_{B3} \\ G_{B4} & G_{B2} \end{smallmatrix}\bigr)$. $\tilde{R}_1^{n_A \times n_B}$, $\tilde{R}_2^{n_A \times n_B}$, $\tilde{Y}_1^{m \times m}$, and $\tilde{Y}_2^{m \times m}$ have non-zero elements only on the main diagonal by construction. We impose the same structure on all $G_{Ai}^{m \times n_A}$ and $G_{Bi}^{m \times n_B}$, for $i=1,2,3,4$. Moreover, we let $G_{A3}=G_{A4}=\mathbf{0}$. This choice of structure for $G_A$ and $G_B$ reduces the number of free parameters in the proposed solution of (\ref{DI/FMI parameter relation 3}) even more. However, it facilitates its solution while leaving plenty of free parameters. Then, (\ref{DI/FMI parameter relation 4}) takes the following form:
\begin{eqnarray*}
2\left(\begin{array}{cc}
\tilde{R}_1 & 0 \\
0 & \tilde{R}_2 \\
\end{array}\right)=
\left(\begin{array}{cc}
G_{A1}^{\top} & 0 \\
0 & G_{A2}^{\top} \\
\end{array}\right)
\left(\begin{array}{cc}
\tilde{Y}_1 & I \\
-I & \tilde{Y}_2 \\
\end{array}\right)
\left(\begin{array}{cc}
G_{B1} & G_{B3} \\
G_{B4} & G_{B2} \\
\end{array}\right),
\end{eqnarray*}
which results into the following set of equations:
\begin{eqnarray}
2\tilde{R}_1 &=& G_{A1}^{\top}(\tilde{Y}_1 G_{B1}+G_{B4}), \label{DI/FMI parameter relation 5} \\
0 &=& G_{A1}^{\top}(\tilde{Y}_1 G_{B3}+G_{B2}), \label{DI/FMI parameter relation 6} \\
0 &=& G_{A2}^{\top}(\tilde{Y}_2 G_{B4}-G_{B1}), \label{DI/FMI parameter relation 7} \\
2\tilde{R}_2 &=& G_{A2}^{\top}(\tilde{Y}_2 G_{B2}-G_{B3}). \label{DI/FMI parameter relation 8}
\end{eqnarray}
Equations (\ref{DI/FMI parameter relation 6}) and (\ref{DI/FMI parameter relation 7})  are satisfied if we choose
\begin{eqnarray}
G_{B2} &=&-\tilde{Y}_1 G_{B3},\ \mathrm{and} \label{DI/FMI parameter relation 9} \\
G_{B1} &=& \tilde{Y}_2 G_{B4}, \label{DI/FMI parameter relation 10}
\end{eqnarray}
respectively. Substituting $G_{B1}$ from (\ref{DI/FMI parameter relation 10}) into (\ref{DI/FMI parameter relation 5}), and $G_{B2}$ from (\ref{DI/FMI parameter relation 9}) into (\ref{DI/FMI parameter relation 8}), we end up with
\begin{eqnarray}
2\tilde{R}_1 &=& G_{A1}^{\top}(\tilde{Y}_1 \tilde{Y}_2 +I)\, G_{B4}, \label{DI/FMI parameter relation 11} \\
2\tilde{R}_2 &=& -G_{A2}^{\top}(\tilde{Y}_2\tilde{Y}_1 +I)\, G_{B3}. \label{DI/FMI parameter relation 12}
\end{eqnarray}
The above matrix equations are equivalent to the following set of scalar equations:
\begin{eqnarray}
2\tilde{R}_{1,ii} &=& G_{A1,ii}(\tilde{Y}_{1,ii} \tilde{Y}_{2,ii} +1)\, G_{B4,ii}, \label{DI/FMI parameter relation 13} \\
2\tilde{R}_{2,ii} &=& -G_{A2,ii}(\tilde{Y}_{1,ii} \tilde{Y}_{2,ii} +1)\, G_{B3,ii},\, i=1,\ldots,i_{max}=\min\{n_A,n_B\}, \label{DI/FMI parameter relation 14}
\end{eqnarray}
with the understanding that for any matrix $T$ in these expressions, $T_{ii}=0$, if $i$ exceeds its row or column dimension. The maximum value of $i$ for the matrices $\tilde{R}_1$ and $\tilde{R}_2$, $i_{max}$, is the minimum of the dimensions of $\tilde{R}_1$ and $\tilde{R}_2$, $\min\{n_A,n_B\}$. We see that there is no need for $m$ to be larger than $\min\{n_A,n_B\}$. On the other hand, $m$ cannot be too small because that would force a number of not necessarily zero last diagonal elements of $\tilde{R}_1$ and $\tilde{R}_2$ in (\ref{DI/FMI parameter relation 13}) and (\ref{DI/FMI parameter relation 14}), to become equal to zero. By way of construction, $\tilde{R}_1$ and $\tilde{R}_2$ have at most $\lceil\frac{\Rank(R_{AB})}{2}\rceil$ non-zero elements on the main diagonal, hence $m$ should be greater or equal to $ \lceil\frac{\Rank(R_{AB})}{2}\rceil$. It is straightforward to see that (\ref{DI/FMI parameter relation 13}) and (\ref{DI/FMI parameter relation 14}) can be satisfied by a multidimensional family of parameters $G_{A1,ii}$, $G_{A2,ii}$, $G_{B4,ii}$, $G_{B3,ii}$, $\tilde{Y}_{1,ii}$, and $\tilde{Y}_{2,ii}$. Hence, there exists a multidimensional family of real $2m \times 2n_A$ and $2m \times 2n_B$ matrices $C_A$ and $C_B$, respectively, and real $2m \times 2m$ $\sharp$-skew-symmetric matrix $X$ satisfying (\ref{eq:DI/FMI parameter relation}), as long as $m \geq \lceil\frac{1}{2} \Rank(\bar{R}_{AB})\rceil$.$\blacksquare$

\section{An Illustrative Example}
\label{sec:An Illustrative Example}

Consider the Hamiltonian interaction between a two degree-of-freedom LQSS $\bar{A}$, and a three degree-of-freedom LQSS $\bar{B}$, with
\[ \bar{R}_{AB}=
\left(\begin{array}{crcrcc}
4&-7&7&0&2&0\\
1&-5&5&-4&1&5\\
9&-6&0&0&2&9\\
12&-8&2&4&3&4\\
\end{array}\right). \]
The special form of the SVD of $\bar{R}_{AB}$ is obtained from the usual SVD as
\[ \bar{R}_{AB}=Q_1 \tilde{R} Q_2^{\top}= Q_1
\left(\begin{array}{cc}
\tilde{R}_1 & 0 \\
0 & \tilde{R}_2
\end{array}\right) Q_2^{\top}, \]
with
\[ \tilde{R}_1 = \left(\begin{array}{ccc}
 22.9090&0&0\\
0&9.2570&0\\
\end{array}\right), \
\tilde{R}_2 = \left(\begin{array}{ccc}
7.4488&0&0\\
0&0&0\\
\end{array}\right), \]
\[ Q_1 =
\left(\begin{array}{rrrr}
-0.3732&-0.6038&0.4961&0.5000\\
-0.2874&-0.6466&-0.4993&-0.5000\\
-0.5793&0.3075&-0.5655&0.5000\\
-0.6652&0.3503&0.4299&-0.5000\\
\end{array}\right), \]
and
\[ Q_2 =
\left(\begin{array}{rrrrrr}
-0.6538&0.4223&-0.1718&0.2086&0.3187&-0.4687\\
0.5608&0.3038&0.0370&-0.1372&0.7537&-0.0705\\
-0.2348&-0.7302&-0.1427&0.2465&0.5395&0.1986\\
-0.0660&0.4308&-0.0490&0.4990&0.0385&0.7465\\
-0.1828&-0.0203&0.9726&0.0875&0.1120&-0.0048\\
-0.4065&0.1010&-0.0197&-0.7876&0.1588&0.4227\\
\end{array}\right). \]
The rank of $\bar{R}_{AB}$ is $3$, hence the minimum number of interconnection channels $m$ is $\lceil \frac{3}{2}\rceil =2$. We will set $m=2$. To satisfy (\ref{DI/FMI parameter relation 11}) - (\ref{DI/FMI parameter relation 12}), we choose $\tilde{Y}_1 = \tilde{Y}_2 = G_{A1} = G_{A2} = I_2$ and, hence, $G_{B4}=\tilde{R}_1$, and $G_{B3}=-\tilde{R}_2$. Then, from (\ref{DI/FMI parameter relation 9}) - (\ref{DI/FMI parameter relation 10}) we have that, $G_{B1}=\tilde{R}_1$, and $G_{B2}=\tilde{R}_2$. Hence, $G_A = I_4$, and
\[ G_B =
\left(\begin{array}{cccccc}
22.9090&0&0&-7.4488&0&0\\
0&9.2570&0&0&0&0\\
22.9090&0&0&7.4488&0&0\\
0&9.2570&0&0&0&0\\
\end{array}\right). \]
For the real orthogonal symplectic matrix $P$, we make the choice $P=I_4$. Then, using the definitions of $G_A$ and $G_B$, $G_A=P C_A Q_1$, and $G_B=P C_B Q_2$, respectively, we compute
\[ C_A =
\left(\begin{array}{rrrr}
-0.3732&-0.2874&-0.5793&-0.6652\\
-0.6039&-0.6466&0.3075&0.3503\\
0.4961&-0.4993&-0.5655&0.4299\\
0.5000&-0.5000&0.5000&-0.5000\\
\end{array}\right), \]
and
\[ C_B =
\left(\begin{array}{rrrrrr}
-16.5307&13.8692&-7.2157&-5.2280&-4.8396&-3.4451\\
3.9092&2.8125&-6.7593&3.9878&-0.1884&0.9354\\
-13.4228&11.8248&-3.5436&2.2052&-3.5366&-15.1783\\
3.9092&2.8125&-6.7593&3.9878&-0.1884&0.9354\\
\end{array}\right). \]
Finally, $Y=P^{\top} \tilde{Y} P=I_4$, and, hence, $X=\mathbb{J}^{-1}Y=-\mathbb{J}$. Then, from (\ref{eq:Static Interconnection Gain}) we have that
\begin{eqnarray*}
\Sigma = (X-I)(X+I)^{-1} = -(I_4 + \mathbb{J}_4)(I_4 - \mathbb{J}_4)^{-1} = -\mathbb{J}.
\end{eqnarray*}
Given any Hamiltonian matrices $\bar{R}_A$ and $\bar{R}_B$ for the LQSSs $\bar{A}$ and $\bar{B}$, respectively, the corresponding Hamiltonian matrices $R_A$ and $R_B$ of LQSSs $A$ and $B$ can be computed by (\ref{eq:New Hamiltonian system A}) - (\ref{eq:New Hamiltonian system B}). As mentioned in the introduction, this method can work even in the case where the systems $\bar{A}$ and $\bar{B}$ are non-linear, provided that a) the additional linear dynamics generated by the Hamiltonians $-\frac{1}{4}x_A^{\top}\mathbb{J}(C_A^{\sharp} X C_A)x_A$, and $-\frac{1}{4}x_B^{\top}\mathbb{J}(C_B^{\sharp} X C_B)x_B$ can be realized, see (\ref{eq:New Hamiltonian system A}) - (\ref{eq:New Hamiltonian system B}), and b) the additional (linear) inputs/outputs can be created in the corresponding systems.

\bibliographystyle{ieeetr}
\bibliography{C:/Users/Symeon/Documents/AAA/Work/Latex/MyBibliographies/Linear_Quantum_Stochastic_Systems,C:/Users/Symeon/Documents/AAA/Work/Latex/MyBibliographies/Books,C:/Users/Symeon/Documents/AAA/Work/Latex/MyBibliographies/My_papers,C:/Users/Symeon/Documents/AAA/Work/Latex/MyBibliographies/Miscellaneous_papers}
\end{document}